# Transformations of Liquid Metals in Ionic Liquid

Fujun Liu[1], Yongze Yu[1], Jing Liu[1, 2] *

1 Beijing Key Lab of Cryo-Biomedical Engineering and Key Lab of Cryogenics, Technical Institute of Physics and Chemistry, Chinese Academy of Sciences, Beijing 100190 , China

2 Department of Biomedical Engineering, School of Medicine, Tsinghua University, Beijing 100084 , China

* To whom correspondence should be addressed: Dr. Jing Liu. Technical Institute of Physics and Chemistry, Chinese Academy of Sciences, Beijing 100190 , China. Tel.: +86-10-82543765. Fax: +86-10-82543767. E-mail: jliu@mail.ipc.ac.cn

**Abstract:** Experimental studies were carried out on the motions and transformations of liquid metal in ionic liquid under applied electric field. The induced vortex rings and flows of ionic liquid were determined via the photographs taken sequentially over the experiments. The polarization of electric double layer of liquid metals was employed to explain the flow of ionic liquid with the presence of liquid metal. Unlike former observation of liquid metal machine in conventional solution, no gas bubble was generated throughout the whole experiments, which makes it possible to encapsulate the liquid system for fabricating soft robots.

**1 Introduction**

According to some medical and military demands, the fabrication and development of soft robots, that contain no or few rigid internal structural elements, is causing increasing concern from both science and engineering categories. Normally soft robots are loosely modeled into animals with non-rigid body parts, such as starfish, squid, and others [1-4]. Compared with traditional 'hard' robots, soft robots possess several outstanding benefits such as that a soft body enables the robot to traverse small openings and reconstitute shape, and survives from large impact force on falling. Moreover, the peristaltic locomotion of soft robots is slow but stable, with reduced noise generation and less requirement of form-factor than legged or wheeled locomotion. In order to achieve delicate tasks, rough terrain negotiation, recovery from overturning, and safety in human interaction, many of the existing soft robots are actuated pneumatically using gas transferred to them from a stationary source via a flexible tether [5-8]. To date, pneumatic robots have not been sufficiently large (less than 15cm in dimension), nor actuated at sufficiently high pressures to support bigger size or weight of commercially available compressed gas cylinders and the other components necessary to operate autonomously.

As a kind of newly emerging functional material, gallium-based liquid metal, which assumes the liquid state at room temperature, shows many interesting and unexpected properties and phenomena. Since the phenomenon of diverse transformations was reported [9], more and more properties and potent applications of liquid metal are studied, including 3D printing, medical imaging and so on [10-15]. When immerged in electrolyte solution, the shape and pattern of liquid metal can be controlled by an applied electric field, like transforming from a large film into a sphere or desired flat shapes with sharp angles [12, 16]. The transformations of liquid metal were due to the change of surface tension caused by electrowetting, which can be described by the Young-Lippmann equation:

$$\cos\theta = \cos\theta_0 + \frac{\varepsilon_0 \varepsilon_r V^2}{2d\gamma} \qquad (1)$$

where $\theta$ and $\theta_0$ are the contact angles with and without applied electric field, respectively; $\varepsilon_0$ is the permittivity of vacuum; $\varepsilon_r$ is the dielectric constant of the insulating material; $d$ is the dielectric thickness and $\gamma$ is the interfacial tension of liquid metal in contact with the ambient phase (air or liquid) [17].

According to Equation 1, the contact angle of liquid metal will decrease with increasing the applied potential, resulting in transformation into sphere with the biggest volume surface ratio and reduced interfacial tension. When the applied potential gradually increased, the formed liquid metal sphere acted as a powerful pump to drive liquid to circulate in designed channel [10], which was explained by an electric double layer (EDL) theory, that a potential gradient will arise under the applied electric field and the charge distribution on the surface of the liquid metal droplet is not uniform, which leads to different pressures between the electrolyte and liquid metal droplet and initiates the mechanical movement of liquid. Many social media and researchers value this phenomenon as a potential power supplement for soft robots, due to the nonuse of compressed gas cylinders or electromotor. However, the electrolysis of electrolyte solution (generally NaOH or KOH solutions in recent reports) cannot be ignored that generated hydrogen bubbles made it pretty hard to encapsulate the pump system as formed high pressure will explode the soft robot.

Therefore, the development of non-electrolysis solution is the key to apply liquid metal as the power supply of soft robots.

The room-temperature ionic liquids (RTIL), an increasingly important set of electrolytes, are organic salts of liquid state at room temperature (generally below 100$^o$C) [18-21], with unique physicochemical characteristics including no significant vapor pressure, non-flammability, good thermal stability and a wide usable temperature range. They are very stable with high ionic conductivity and therefore have promising prospects in electrochemical applications like lithium ion batteries [22-24]. In the same way, RTIL can be used as the electrolyte for liquid metal pump without any electrolysis of solution or generation of hydrogen gas. Therefore in this paper, we investigated the motion of ionic liquid driven by liquid metal droplets under applied electric field, and tested the transformation behaviors of liquid metal. To the best of our knowledge, this is the first ever demonstration of liquid metal pump in ionic liquid, and the results will contribute to the design of new-style soft robots.

## 2 Materials and Methods

Glass slides (Sail Brand, YanchengXingfu Glass Instrument Factory, Yancheng, China) were cleaned with isopropanol and dried in a stream of filtered nitrogen. To achieve good record of liquid movements, the cleaned glass slide was suspended on brightness-controllable LED light. For simplifying the experimental variables to consider, pure gallium (99.999 %, Anhui Rare New Materials Co., Anhui, China) was adopted instead of GaIn-based alloy as widely used in previous reports [9, 12, 16].1-butyl-3-methylimidazoliumtetrafluoroborate (99%, Chengjie Chemical LLC., Shanghai, China) was taken as the electrolyte liquid, a direct-current source was used as the potential supply and a digital camera Canon 70D was used to record the experimental phenomena. The equipment was presented in Fig. 1-a.

## 3 Results and Discussions

### 3.1 Influence of liquid metal on electro-induced shrink of ionic liquid

To fully characterize our experimental system, we firstly studied the electro-induced shrink of ionic liquid. When a direct-current electric field was applied to the ionic liquid, a wavelet diffused from the negative electrode all around with low applied potential (3.5 V). If increasing the applied potential to 12 V, the shrink of ionic liquid got severer until a "hole" appeared that ionic liquid tided from the negative electrode and recovered subsequently like ocean waves, as shown in Fig. 1-(b-f). The frequency of this "spreading-retraction" shrink was 0.1 Hz and accelerated to 2 Hz when a droplet of liquid metal was immersed in ionic liquid. Moreover, an advection flow run from positive to negative electrode, bypassing the liquid metal droplet and forming two vortices near the boundary plane of negative electrode (the flow direction was shown with white dashed line in Fig. 1-g).

Ionic liquids were always studied due to their applications as clean solvents and electrochemical electrolyte, but the shrink of ionic liquid has never been reported. The electrowetting of ionic liquid droplets was studied and the results showed that the ionic liquid droplet transformed from sphere to flat shape, due to the decrease of contact angle under applied voltage [25]. According to the model in that report, the ionic liquid should be at a stable state, which is completely conflicting with our experimental results shown in Fig. 1-(b-f). Therefore, we made a hydrodynamic model to explain the "spreading-retraction" shrink of ionic liquid. The used ionic liquid BMT (1-butyl-3-methylimidazolium tetrafluoroborate) was comprised by the anion $BF_4^-$ and the cation $C_8H_{15}N_2^+$, which will be polarized under external voltage that negative and positive charges assembled near the negative and positive electrodes, respectively, as shown in Fig. 2-a. Due to the small formula weight and ignorable stereo-specific blockade, the anion $BF_4^-$ was much easier to migrate than the cation $C_8H_{15}N_2^+$. Therefore, as the function of the electric field force $F_E$, ionic liquid nearby the negative electrode spreads to all around radically, as shown in Fig. 2-b. The field force $F_E$ was identified by yellow arrow, and followed the equation:

$$F_E = qE = qU_{AB}/d \tag{2}$$

where q is the quantity of electric charge, E is the electric field, $U_{AB}$ is the voltage between two points A and B, and d is the distance between A and B on the electric field direction. With aggravation of the spreading and insulation of ionic liquid from the negative electrode, the negative and positive charges will neutralize (q = 0 and $F_E = 0$) to achieve the equi-distribution of electric charges in ionic liquid, as shown in Fig. 2-c. Then the force of gravity *G* played a leading role, leading to the retraction of ionic liquid to get in touch with the negative electrode again, as shown in Fig. 2-d, and the processes of Fig. 2-(a-d) repeated periodically.

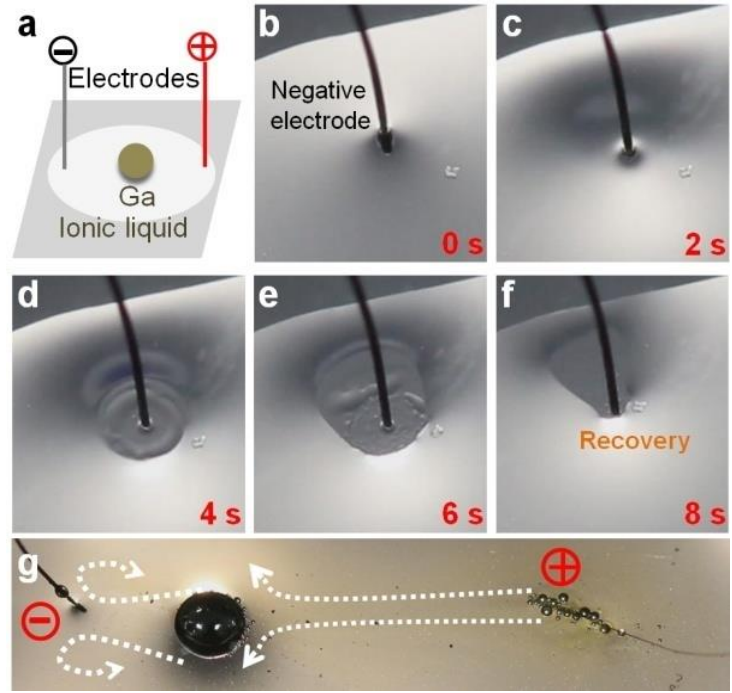

**Fig. 1** Schematics for the experiments to study the transformation of IL-LM (ionic liquid and liquid metal) system (a), the electro-induced shrink phenomenon of ionic liquid (b-f) with a complete circle, and the ionic liquid flow induced by liquid metal droplet (d).

The "spreading-retraction" shrink of ionic liquid had a prominent change with the immersion of liquid metal droplet, that the frequency of shrink increased about 20 times and smooth flow run from the positive electrode to the negative one bypassing the liquid metal droplet, as shown in Fig. 1-g. Due to the special fluidity and lattice structure, negative charges gather on the liquid metal surface, resulting in an accumulation of positive charges in a diffuse layer of electric double layer (EDL), as shown in Fig. 2-e. Without external electric field, the negative and positive charges equally disturb on the surface and EDL, respectively. Under an external voltage, the liquid metal droplet was polarized that negative charges gathered to the direction of positive electrode and positive charges gathered to the direction of negative electrode, as shown in Fig. 2-f. The existence of polarized liquid metal made great effect on the spreading of ionic liquid, by decreasing d in Equation 2 from the distance between the two electrodes to that between the negative electrode and liquid metal, because the polarized liquid metal acted as a new electrode. Thus the polarized ionic liquid was easier and quicker to achieve the equi-distribution of electric charges during each loop of the "spreading-retraction" shrink, leading to the prominent increase of shrink frequency. Another influence of liquid metal was the induced flows from the positive electrode to liquid metal and from liquid metal to the negative electrode. After the polarization, the surface and EDL of liquid metal droplet acted as a charged capacitor, leading to the change of surface tension between the liquid metal and ionic liquid according with Equation 1. So it induced two pressure differences, the one between the positive electrode and liquid metal $\Delta p_{Pos-LM}$ and the other one between liquid metal and the negative electrode $\Delta p_{LM-Neg}$, due to different surface tensions at each position. Under the external voltage, the pressure difference continuously drives ionic liquid to flow from high-surface-tension part to low-surface-tension part like a pump, as shown in Fig. 2-f.

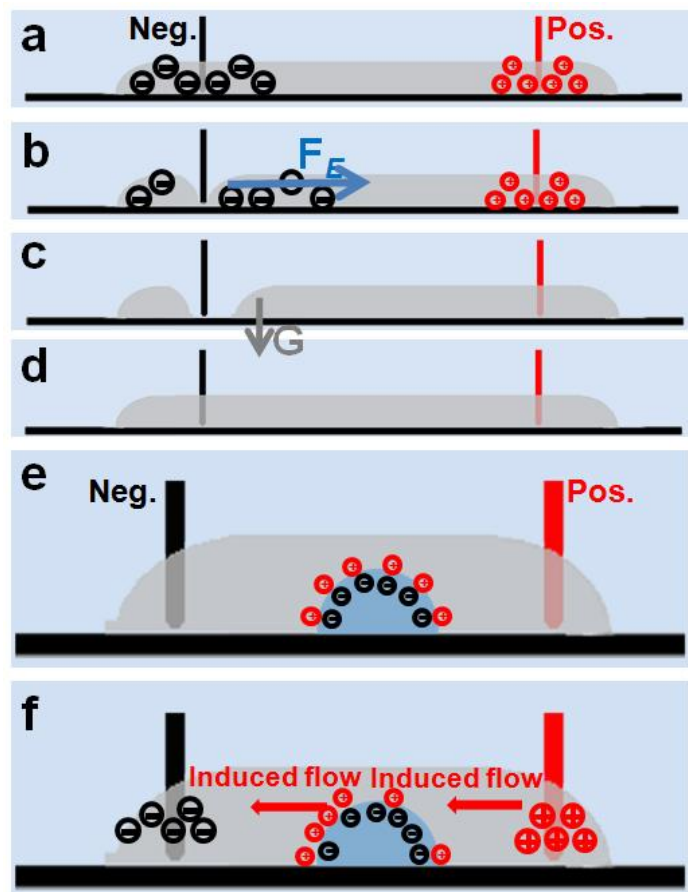

**Fig. 2** Schematic of "spreading-retraction" shrink of ionic liquid (a-d), and the influence of liquid metal droplet on the shrink of ionic liquid (e, f). Pos. and Neg. represent the positive and negative electrodes, respectively.

**3.2 Turbulent flow and vortex of multi-droplet system**

To further clarify the influences of liquid metal droplet on the induced flow of ionic liquid and verify the above model, multiple liquid metal droplets (number = 2-5) were immersed in ionic liquid to investigate the induced flow and vortex. The results were shown in Fig. 3. When two liquid metal droplets were placed along the electric field, flow of ionic liquid run between the two droplets with the same direction as the electric field, two vortexes appeared on the inside of the left droplet, and the left droplet moved right, as indicated in Fig. 3-a. As the two droplets were arranged perpendicular to the electric field, two flows run from the positive electrode to two droplets, respectively, two vortexes appeared on the right side of each droplet and no droplet showed any movement, as indicated by Fig. 3-b. When three droplets were arranged perpendicular to the electric field, two flows run from the positive electrode through the gaps between droplets, two vortexes appeared on the external right sides of the upper and lower droplets as shown in Fig. 3-c, and both of the upper and lower droplets moved to the middle one. As the number of droplets increased to four, the induced flow and movements of droplets depended on how droplets arranged, as shown in Fig. 3-d and Fig. 3-e, respectively. An interesting phenomenon appeared with five droplets, that the only movement was the migration of the right droplet towards the positive electrode, as shown in Fig. 3-f.

The induced flow was closely dependent on the polarization of liquid, not only the overall charge distribution but a complicated combination and mutual influence of each position of liquid metal and ionic liquid. By comparing Fig. 3-a and Fig. 3-b, it was clear that the polarization occurred along the electric field, as the flow runs from the positive side of right droplet to the negative side of left droplet in Fig. 3-a and from the positive electrode to the negative sides of droplets in Fig. 3-b. The vortex can invert to advection flow when the third droplet was placed to decrease the gap between interfacing droplets, as shown in Fig. 3-c. According to the previous report [26], the radius of vortex ring $a$ of an ideal fluid changes according the following equation:

$$a(\log\frac{8a}{\varepsilon}-\frac{7}{4})-2a\csc\theta(F-E)+a\sin\theta F = a_0(\log\frac{8a_0}{\varepsilon_0}-\frac{7}{4}) \quad (3)$$

where $a_0$ and $\varepsilon_0$ are the radius of the vortex ring and its core, respectively, $\theta$ is a half angle which the vortex subtends at the boundary plane, $F$ and $E$ are the complete elliptic integrals of the first and second kind, respectively. When two induced vortexes crossover, and the radius of the vortex core $a_0$ increased until bigger than the radius of the vortex ring $\varepsilon_0$, leading to that the Equation 3 will not be applicable to this situation. Then the vortexes changed to advection flows through the gaps in Fig. 3-c. This theory can also explain the phenomena of four-droplet and five-droplet systems.

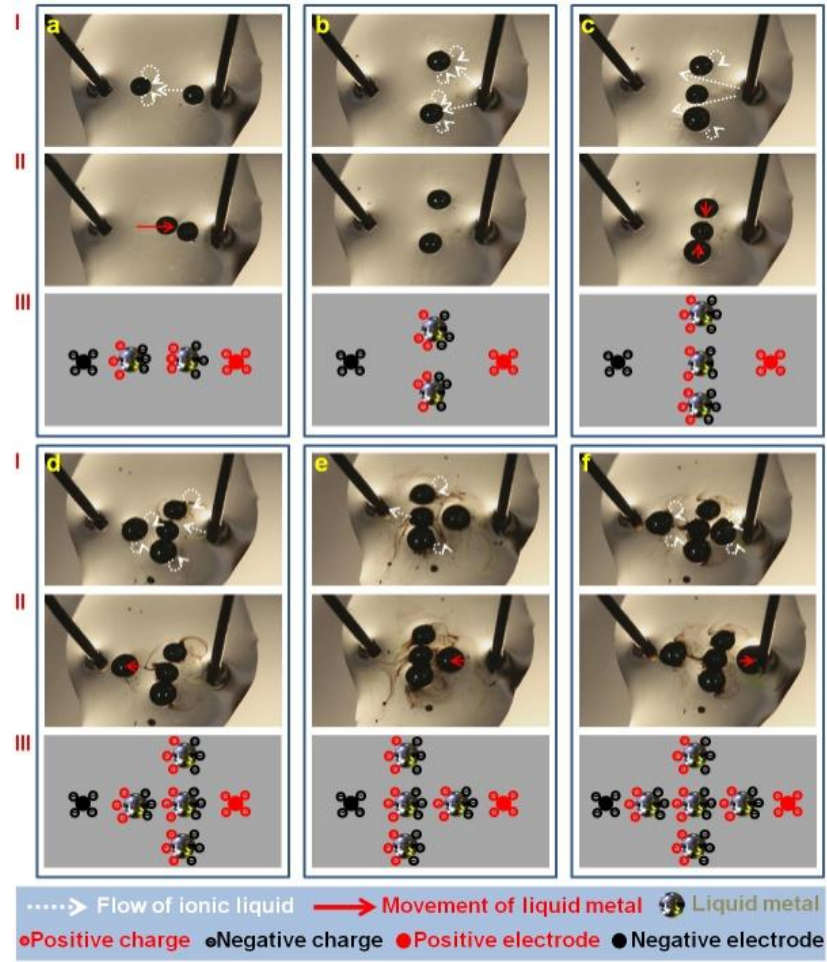

**Fig. 3** The flow of multi-droplet system under applied electric field. The white dashed line in (I) image of each blue frame indicates the direction of induced flow of ionic liquid, the red line in (II) image indicates the movements of the liquid metal droplets and the schema of charge distribution was described in (III) image. The left and right electrodes were negative and positive, respectively, in each image.

**3.3 The transformation of liquid metal in ionic liquid**

Finally, we consider the transformation of liquid metal in ionic liquid. A strip of liquid metal was immersed in the ionic liquid, between the positive and negative electrodes. The distance between two electrodes and the length of liquid metal were 1 cm and 0.9 cm, respectively. As shown in Fig. 4-a, the constriction of liquid metal happened when an external voltage of 5 V was applied. The shape of liquid metal transformed from original strip shape into a sphere. The speed of the constriction was also recorded, from 0.02 cm/s at the beginning to 0.002 cm/s after 100s, as shown in Fig. 4-b.

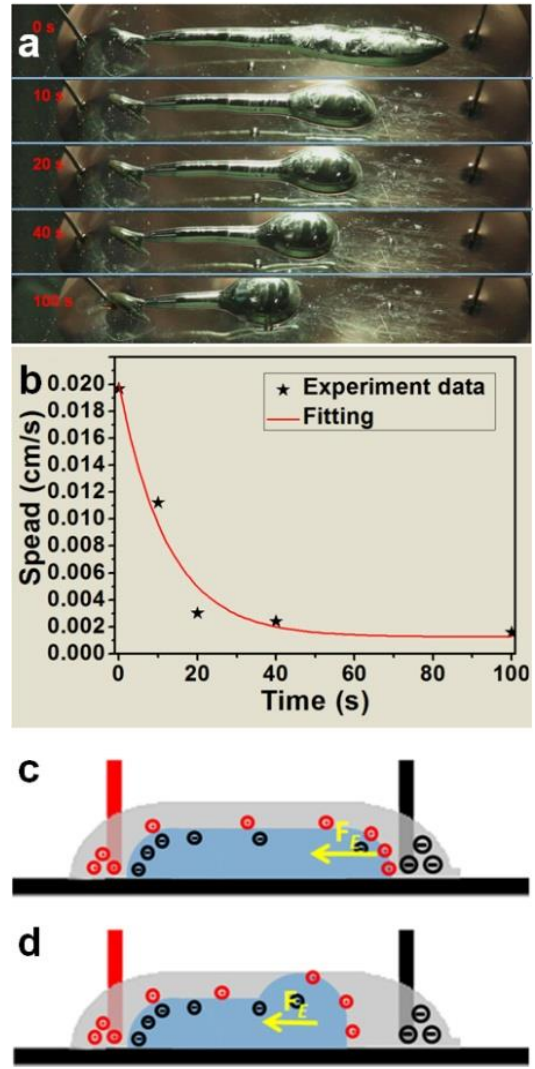

**Fig. 4** a) The transformation of liquid metal in ionic liquid. b) indicated the speed of constriction of liquid metal from a strip shape to a sphere, where the left electrode was positive and the right one was negative. c) and d) described the electric charge distribution of the liquid system and the force analysis of liquid metal before and after constriction, respectively.

The principle of polarization of liquid metal was described above, where the "electric double layer" theory played an important role. Due to its metal line property, the negative charges of liquid metal are composed by free electrons, and the positive charges are from metal ions. The movement of polarized liquid metal occurred on the positive part. Therefore only one part of the liquid metal near the negative electrode moved along the electric field direction as shown in Fig. 4-a. Fig. 4-c graphically represented the force analysis of polarized liquid metal under external electric field. Due to Equation 2, the electric field force $F_E$ depended on the quantity of positive charge $q$, as the electric field $E$ was constant. Accompanied with the constriction, the electric charge distribution of liquid metal got more and more equal, leading to the decrease of electric field force and speed of the movement, as shown in Fig. 4-d. Finally liquid metal transformed into the shape of sphere to minimize the superficial area, against the increase of surface tension led by the external electric field.

In the other hand, the transformation of liquid metal also can be studied in a surface tension perspective. The surface tension between the liquid metal and ionic liquid greatly depends on the capacitance of EDL and the potential difference, according to the Lippman's equation,

$$\gamma_V = \gamma_0 - \frac{1}{2}CV^2 \qquad (4)$$

where $\gamma_V$, $C$ and $V$ are the surface tension, the capacitance and the potential difference across the electrical double layer, respectively, and $\gamma_0$ is the maximum surface tension when $V=0$ [27]. Therefore, the variety of surface tension shows linear tendency with the external potential difference. According to previous report [28], the moving and transformation of liquid metal was mostly driven by the pressure jump $p$ across the interface between the liquid metal and ionic liquid, which can be described by this equation,

$$p = \gamma_V(2/r) \qquad (5)$$

where $r$ is the apparent radius of the liquid metal droplet. A combination of Equation 4 and Equation 5 led to a conclusion that the surface tension $\gamma_V$ can affect the pressure difference $p$ between liquid metal and ionic liquid. The pressure difference $p$ acted as the driving force to change the shape of liquid metal from original strip shape into a sphere. With the assembling of liquid metal, the apparent radius $r$ increased to give rise to the decrease of the pressure difference $p$, therefore the constriction speed of liquid metal decreased evidently as shown in Fig. 4-b.

**4 Conclusion**

In summary, the induced flow of ionic liquid induced by liquid metal was discovered. To investigate the potential application of liquid metal as pumps, multiple droplets were employed to evaluate the induced advection flow and vortex. The results showed that it was the distribution of electric charges that significantly influences the dynamic change of the liquid system. In addition, electric-field-induced transformations of liquid were also investigated. Compared with previous reports, this study offered feasible opportunities for the power supply of soft robots due to avoidance of the gas bubbles during the experiments. With finely designed electric circle, smart packaged robots with multi-dimension variance can possibly be fabricated in the coming time.

**Acknowledgments** This work was partially supported by Joint Funds of Ministry of Education, the Dean's Research Funding and Frontier Project from Chinese Academy of Sciences.

**Conflict of interest** The authors declare that they have no conflict of interest.